\documentclass[aps,prb,reprint,groupedaddress,showpacs]{revtex4-1}

\usepackage{bm}

\begin{document}


\title{Spin Sensitive Electron Transmission Through Helical Potentials}


\author{A. A. Eremko and V. M. Loktev}
\email[]{eremko@bitp.kiev.ua; vloktev@bitp.kiev.ua}
\thanks{}
\affiliation{Bogolubov Institute for Theoretical Physics, Kyiv, Ukraine}


\date{\today}

\begin{abstract}
We calculate the transmission coefficient for electrons passing through the helically shaped potential barrier which can be, for example, produced by DNA molecules.
\end{abstract}


\maketitle

\emph{Introduction}. In recent years, it was discovered that electron transmission through ordered thin films of chiral molecules is highly spin selective \cite{Scie99, PRL06, MRSBul10, Sci11, NanoLet11}. This effect was termed as chiral-induced spin selectivity (CISS) effect \cite{Sci11} (for details see review \cite{JPhysChemL12} and Refs. therein). In experiments \cite{PRL06, Sci11}, the transmission of photoelectrons through self-assembled monolayers (SAMs) of double-stranded DNA (dsDNA) on gold has been studied. The spin polarization (SP) of electrons ejected from Au substrate and transmitted through SAM of dsDNA was measured and the strong SP, which is defined as \cite{Sci11} $ P = (I_{\uparrow} - I_{\downarrow})/(I_{\uparrow} + I_{\downarrow}) $ was observed. Here $ I_{\uparrow} $ and $ I_{\downarrow} $ are the intensities of the signals corresponding to the SP oriented parallel and antiparallel to the electrons' velocity, respectively. 

Recently, different models have been proposed \cite{PRL08, JChemP09, PRB12, GuoSun12} to explain experimental results. A scattering theory in the first Born approximation has been applied to obtain the SP in the differential cross section of electrons moving through chiral molecules with energies above the vacuum level\cite{JChemP09}.
The model of point charges placed along a helical line is considered in a tight binding approximation for electronic structure of the helix and the transmission of distinct electron spin state is computed by the Landauer formulation\cite{PRB12}. The SP conductance through a metal-DNA-metal structure is calculated in a tight binding picture\cite{GuoSun12}. Although all these studies differ in details, they possess similar physical basis. Each of them is based on accounting of the spin-orbit interaction (SOI) of an electron that is moving through a helical potential. In other words, describing different aspects of the problem, approaches are based on the Schr\"{o}dinger equation \cite{Dav} 
\begin{equation} 
\label{SEq}
\left[ -\frac{\hbar^{2}}{2m}\triangle + V - i \frac{\hbar^{2}}{4m^{2}c^{2}}  \left(\bm{\sigma} \times \bm{\nabla} V \right)\cdot \bm{\nabla} \right]  \Psi = E \Psi ,
\end{equation}
which describes electron states in a potential $ V \equiv V(\mathbf{r}) $. Here $ \Psi(\mathbf{r} ) $ is an electron spinor wavefunction, $ m $ is the electron mass, and the last term in the brackets describes the SOI: $ H_{SO} = \alpha \left(\bm{\sigma} \times \bm{\nabla } V\right)\cdot \mathbf{p} $ where $ \bm{\sigma } $ is a vector whose components are the Pauli matrices $ \sigma_{j} $ $ (j=x,y,z) $, $ \alpha = \hbar / (2mc)^{2} $ and $ \mathbf{p} = -i \hbar \bm{\nabla} $ is a momentum operator.

In experiments \cite{PRL06, Sci11}, the photoelectrons have an energy above vacuum level as they transmit through the SAM of dsDNA molecules and then to the detector. These experimental results are our main motivation to consider the problem of an electron transmission through the plane-parallel region with the potential $ V(\mathbf{r}) $ caused by ordered dsDNA molecules in a monolayer. Although our sdudy has common physics with above papers \cite{JChemP09, PRB12, GuoSun12}, it differs by the general statement of the problem, and gives complementary features to the SOI role for the CISS. In particular, we calculate analytically continuous energy eigenfunctions of Eq.(\ref{SEq}) to obtain the transmission coefficient for electrons passing through the chiral potential barrier. 

\emph{Potential with chiral symmetry}. In the space, we can, by convention, separate a cylindrical volume with one dsDNA molecule and with an axis which coincides with a molecule's symmetry axis. There the electrostatic field arisen from the charge distribution in a molecule, and hence the potential $ V_{\mathit{mol}} (\mathbf{r}) $ for external electrons, has the very same symmetry as a molecule. The main characteristic of helix-shaped molecules, in general, and dsDNA, in particular, is its helical symmetry which is a requirement for CISS to occure \cite{JPhysChemL12}. Below in Eq.(\ref{SEq}) instead of $ V_{\mathit{mol}} (\mathbf{r}) $, we will use the potential $ V(\mathbf{r}) $ which is the averaged value of $ V_{\mathit{mol}} (\mathbf{r}) $ over atomic structure of a cell with one repeating unit composing a macromolecule. This averaged potential is invariant under continuous helical translation defined as the translation along $ z $ by the distance $ b\tau $ with simultaneous rotation around the $ z $ axis by angle $ 2\pi \tau $ with $ b $ being the pitch of helix-shaped molecule and $ \tau $ continuous dimensionless parameter. Under such translation a point with coordinate $ (x,y,z) $ moves along the helical curve represented in a parametrized form 
\begin{equation}
\label{vint}
\mathbf{r}(\tau) = \rho \cos \left( \varphi \pm 2\pi\tau \right) \mathbf{e}_{x} + \rho \sin \left(\varphi \pm 2\pi\tau \right) \mathbf{e}_{y} + \left( z + b \tau \right) \mathbf{e}_{z} ,
\end{equation}
where $ \rho = \sqrt{x^{2} + y^{2}} $, $ \varphi = \arctan (y/x) $, and the sign ``$ + $'' corresponds to a right- and ``$ - $'' to a left-handed helix. 

The averaged potential $ V(\mathbf{r}) $ is invariant under continuous helical translation and hence the curves (\ref{vint}) are the equipotential lines. Therefore the equality $ V\left( x(\tau),y(\tau),z(\tau)\right) - V\left( x,y,z\right) = 0 $ takes place for arbitrary value of the parameter $ \tau $. For infinitesimal helical displacements this equality leads to the differential relation which in coordinates $ (\rho, \varphi, z) $ can be written as
\[ \pm \frac{\partial V}{\partial \varphi} + \frac{b}{2\pi} \frac{\partial V}{\partial z} = 0 . \]
From this relation one can find that the helical potential is characterized by the following dependance on cylindrical coordinates
\begin{equation}
\label{difV2}
V(\rho,\varphi,z) = V(\rho,\varphi \mp\ \frac{2\pi}{b} z )
\end{equation} 
with $ - / + $ for right-/left-handed helical symmetry. Below we will consider helicies with the right-handed symmetry corresponding to B-DNA molecules. 

Each curve is characterized by a Frenet frame -- a moving reference frame of three orthonormal vectors $ \mathbf{e}_{j} $ which describes a curve locally at each point. For the helical curve (\ref{vint}), the Frenet frame is 
\begin{eqnarray}
\label{n,t,b}
\mathbf{e}_{t} = -\frac{2\pi\rho \sin \varphi}{a(\rho)} \mathbf{e}_{x} + 
\frac{2\pi\rho \cos  \varphi}{a(\rho)} \mathbf{e}_{y}  + \frac{b}{a(\rho)} \mathbf{e}_{z} , \nonumber \\
\mathbf{e}_{b} = \frac{b \sin \varphi}{a(\rho)} \mathbf{e}_{x} - \frac{b \cos  \varphi}{a(\rho)} \mathbf{e}_{y} + 
\frac{2\pi\rho}{a(\rho)} \mathbf{e}_{z} ,  \\ 
\mathbf{e}_{\rho} = \cos \varphi \mathbf{e}_{x} + \sin \varphi \mathbf{e}_{y}  .\nonumber
\end{eqnarray}
where $ a(\rho) = \sqrt{b^{2} + (2\pi \rho)^{2}} $ is the arc length of one turn of the helix of pitch $ b $ and radius $ \rho $. The orthonormal vectors  $ \mathbf{e}_{t} $, $ \mathbf{e}_{\rho} $, and $ \mathbf{e}_{b} $ determine, respectively, the tangent, normal, and binormal directions for the equipotential helical curve at the point $ (x,y,z) $.

Calculation $ \bm{\nabla } V $ with taking into account (\ref{difV2}) and (\ref{n,t,b}) gives $ \bm{\nabla}V = V_{\rho} \mathbf{e}_{\rho} + V_{b} \mathbf{e}_{b} $, or 
\begin{equation}
\label{gradV}
\bm{\nabla}V = | \bm{\nabla}V | \mathbf{e}_{V} \quad  \textrm{with} \quad 
\mathbf{e}_{V} = \cos \theta \mathbf{e}_{\rho} + \sin \theta \mathbf{e}_{b} ,
\end{equation}
where $ V_{\rho } = \partial V/\partial \rho $, $ V_{b} = - \left( a(\rho )/b \rho \right) \left( \partial V/ \partial \varphi \right) $, $ | \mathbf{\nabla}V | = \sqrt{V_{\rho}^{2} + V_{b}^{2} } $, and $ \mathbf{e}_{V} $ is a unit vector with $ \tan \theta = V_{b} / V_{\rho} $.

Note that in cylindrical coordinates a parametrized form for right-handed helical lines (\ref{vint}) is 
\begin{equation}
\label{cylpar}
\rho (\tau) = \rho , \quad  \varphi (\tau) = \phi + 2\pi \tau , \quad  z(\tau) = b\tau .
\end{equation}
Here we put $ z(0) = 0 $. Hence the values $ \rho $ and $ \varphi $ indicate the helix which penetrates the plane $ z= 0 $ at these coordinates.

\emph{The electron wave function in a helical potential}. To find a solution of Eq.(\ref{SEq}) let us use the semiclassical, or WKB, approximation \cite{Dav}, i.e. look for a solution 
\begin{equation}
\label{Psi}
\Psi ( \mathbf{r } ) = e^{\frac{i}{\hbar} S} \Phi ,
\end{equation} 
where $ S \equiv S(\mathbf{r}) $ is phase function (action) and $ \Phi \equiv \Phi(\mathbf{r} ) $ is spinor function (probability density amplitude). Substitution (\ref{Psi}) into (\ref{SEq}) results in the set of equations completely equivalent to the Eq.(\ref{SEq}):
\begin{widetext}
\begin{eqnarray}
\label{QCEq}
\left[ \frac{1}{2m} \left( \bm{\nabla} S \right)^{2} - \left( E - V \right) + \frac{\hbar}{4m^{2}c^{2}} \left(\bm{\sigma} \times \bm{\nabla} V\right)\cdot \bm{\nabla} S \right] \Phi = \frac{\hbar^{2}}{2m} \triangle \Phi ,
 \nonumber \\ 
\left( \triangle S\right) \Phi + 2\bm{\nabla} S\cdot\bm{\nabla}  \Phi + \frac{\hbar}{2mc^{2}} \left(\bm{\sigma} \times \bm{\nabla} V\right)\cdot \bm{\nabla} \Phi = 0 .
\end{eqnarray} 
\end{widetext}

In semiclassics the term with $ \hbar^{2} $ in right-hand part of the first equation is neglected and there it is an equation describing the phase fronts of an electron wave. On the analogy with geometrical optics, an electron ray is a line or curve that is ortogonal to wave fronts $ S(\mathbf{r}) = const $ and a $ \bm{\nabla} S $ is proportional to the wavenumber along this line. For the given potential $ V(\mathbf{r}) $ and the fixed energy $ E $ electron trajectories are equipotential helical lines. We ascribe to these lines the direction assuming that the positive direction at each point corresponds to such a direction of $ z $. If a radius-vector $ \mathbf{r}(l) $ of a point which is placed on the line (\ref{vint}) is considered as a function of arc length $ l $ of the ray (natural or arc-length parametrization at which $ \tau = l/a(\rho) $ in (\ref{vint}) and (\ref{cylpar})) then for $ S $ at this point one can write down
\begin{equation}
\label{gradS}
\bm{\nabla} S = \hbar k \frac{d \mathbf{r}(l)}{dl} = \hbar k \mathbf{e}_{t}.
\end{equation}

Taking into eccount (\ref{gradV}) and (\ref{gradS}) in the first equation of (\ref{QCEq}) and neglecting the right-hand part we obtain 
\begin{equation}
\label{SCeq}
\left[ \frac{\hbar^{2} k^{2}}{2m} + V - E - \frac{\hbar^{2} q_{SO} k }{m} \left( \mathbf{e}_{t}\times \mathbf{e}_{V} \right)\cdot \bm{\sigma}  \right] \Phi = 0 ,
\end{equation}
where wavenumber
\begin{equation}
\label{qSO}
q_{SO} = \frac{| \bm{\nabla}V |}{4mc^{2} } 
\end{equation}
characterizes in fact a strength of SOI induced by chiral potential.

Eq.(\ref{SCeq}) has a solution if $ \Phi $ is proportional to the eigenspinor of the matrix $ \left( \mathbf{e}_{t}\times \mathbf{e}_{V} \right)\cdot \bm{\sigma} $ only. Taking into account (\ref{n,t,b}) and (\ref{gradV}) one obtains 
\begin{equation}
\label{spin-eq}
\left( \mathbf{e}_{t}\times \mathbf{e}_{V} \right)\cdot \bm{\sigma} =
\left( \begin{array} {cc}
 -\cos \gamma & e^{-i(\varphi + \phi)}\sin \gamma \\ e^{i(\varphi + \phi)}\sin \gamma & \cos \gamma
\end{array} \right) ,
\end{equation}
where 
\begin{equation}
\label{gamma}
\cos \gamma = \frac{2\pi \rho}{a(\rho)} \cos \theta , \qquad 
\tan \phi = \frac{b}{a(\rho)} \cot \theta .
\end{equation}

Hence $ \Phi = A \chi_{\sigma} $, where normalization multiplyer $ A $ as a function has the same symmetry as the potential $ V $, and $ \chi_{\sigma} = (u_{1} \quad u_{2})^{T} $ is the solution of equation $ \left( \mathbf{e}_{t}\times \mathbf{e}_{V} \right)\cdot \bm{\sigma} \chi_{\sigma} = \sigma \chi_{\sigma} $ with $ \sigma $ being the eigenvalue. The matrix (\ref{spin-eq}) has eigenvalues $ \sigma = \pm 1 $ for two orthonormal spinors 
\begin{equation}
\label{spinors}
\chi_{+} = 
{ e^{-i\frac{\varphi + \phi}{2}} \sin \frac{\gamma}{2} \choose e^{i\frac{\varphi + \phi}{2}} \cos \frac{\gamma}{2} }, \quad 
\chi_{-} = 
{-e^{-i\frac{\varphi + \phi}{2}} \cos \frac{\gamma}{2} \choose e^{i\frac{\varphi + \phi}{2}} \sin \frac{\gamma}{2} } \, .
\end{equation}
Therefore, the SOI defines the spin quantization axis, and explicit form of Eq.(\ref{SCeq}) becomes depending on the electron spin projection:
\begin{equation}
\label{eqE_k} 
\frac{\hbar^{2} k^{2}}{2m} - \sigma \frac{\hbar^{2} q_{SO} k }{m}  + V - E  = 0 .
\end{equation}
This equation gives either the energy dependence upon the wavenumber $ k $ along the ray 
\begin{equation}
\label{E_k}
E_{\pm}\left( k \right) = \frac{\hbar^{2}(k \mp q_{SO})^{2}}{2m} - \frac{\hbar^{2}q_{SO}^{2}}{2m} + V \,,
\end{equation}
or the wavenumber at the given energy 
\begin{equation}
\label{k_E}
k_{\sigma} = \pm \sqrt{ \frac{2m}{\hbar^{2}} \left( E-V \right) + q_{SO}^{2} } + \sigma q_{SO} = 
q + \sigma q_{SO} .
\end{equation}
In Eq.(\ref{k_E}) we have introduced the running wave number 
\begin{equation}
\label{q}
q = \pm \sqrt{\frac{2m}{\hbar^{2}}(E -V) + q_{SO}^{2} }  ,
\end{equation}
which determines the velocity of an electron propagation along helical ray with given energy. Indeed, for velocity we have $ v = (1/\hbar) (dE_{\pm}/dk_{l}) = \hbar (k_{l} \mp q_{SO}/m = \hbar q/m ) $. Signs ``$ \pm $'' in (\ref{q}) correspond to motion in positive and negative directions, respectively. 

Thus, according to the definitions (\ref{gradS}) and (\ref{k_E}) there are two solutions for phase function $ S_{\sigma} $. Bearing in mind the directional derivative of $ S $ along the helical trajectory pointed by the vector $ \mathbf{e}_{t} $ one can see that $ dS/dl = \bm{\nabla} S \cdot \mathbf{e}_{t} = \hbar k $. Because $ k_{\sigma} $ are constant along a trajectory, solutions for $ S_{\sigma} $ have the form 
\begin{equation}
\label{S_sigma}
S_{\sigma} = \hbar k_{\sigma} l + F = \hbar \left( q + \sigma q_{SO} \right) l + F 
\end{equation}
where the integration constant $ F $ has the same symmetry as the potential and can be included into $ A $.

So, the desired functions in the helical potential are 
\begin{equation}
\label{Psi_s}
\Psi_{q,\sigma} = A_{\sigma} e^{i\left(q + \sigma q_{SO} \right)l } \chi_{\sigma} 
\end{equation}
where spinors $ \chi_{\sigma} $ are determined in (\ref{spinors}). These eigenfunctions describe a curvelinear propagation of electrons along helical rays with definite vectors of the SP: $ \mathbf{P}_{\sigma} = \langle \chi_{\sigma} | \bm{\sigma} | \chi_{\sigma} \rangle $, $ \mathbf{P}_{-} = -\mathbf{P}_{+} $. 

Therefore the SOI acting on the moving electron in the helical potential breaks the spin degeneracy. Such a SOI resulting from the lack of inversion symmetry is usually referred to as the Rashba SOI \cite{Rashba}. Due to the last the spin degenerate one-dimensional (along the electron ray) band $ E(k) = \hbar^{2} k^{2} /2m $ splits in the momentum space by $ 2q_{SO} $ into subbands (\ref{E_k}). This is general consequence of a Rashba-like SOI for a quasi-one-dimensional propagation of electrons along curved lines. The Eq.(\ref{E_k}) can be considered as the result of the nearly free electrons approximation. The same conclusion can be obtained also in the tight binding approximation \cite{tbappr}.

In general, the eigenstate of an electron, which propagats in some direction with the energy $ E $, is described by a superposition of the functions (\ref{Psi_s}). This superposition can be written down as $ \Psi_{q}(l) = A \exp (iql) \chi (l)  $ where the spinor $ \chi (l) = \sum_{\sigma}  a_{\sigma} \exp (i\sigma q_{SO}l)\chi_{\sigma} = \left( u_{\uparrow}(l) \quad u_{\downarrow}(l) \right)^{T} $ describes the spin orientation, where $ \sum_{\sigma} |a_{\sigma}|^{2} = 1 $.

So, $ z $-component of the SP vector $ |u_{\uparrow}(l)|^{2} - |u_{\downarrow}(l)|^{2} $ is oscillating function of the variable $ l $ with the wavenumber $ 2q_{SO} $. These oscillations with necessity lead to the CISS effect. To demonstrate this, the electron transmission through the chiral barrier is considered below.

\emph{The transmission coefficient (TC)}. Let the barrier which is due to a SAM of dsDNA, is located between $ z = 0 $ and $ z = d $. It divides the space in three parts: left L ($ z<0 $), central C ($ 0 < z < d $), and right R ($ z > d $) ones. In the parts L and R the potential is constant and can be put zero, meaning that the electron is (quasi)free. In these regions, the phase equals $ S = \hbar \mathbf{k} \cdot \mathbf{r} $ with arbitrary coordinate independent spinor. In the part C, the solution of Eq.(\ref{SEq}) is given by the expression (\ref{Psi_s}). To obtain the TC, consider the certain situation: an electron normally incident upon the barrier from the left side L. In this case the solution of the Schr\"{o}dinger Eq.(\ref{SEq}) can be written as 
\begin{eqnarray}
\label{gensol}
\Psi_{L} = A_{\mathit{inc}} e^{ik_{0}z} \chi_{\mathit{inc}} + A_{\mathit{ref}} e^{-ik_{0}z } \chi_{\mathit{ref}} , \quad z < 0 ,  \nonumber \\ 
\Psi_{C} = \sum_{\nu , \sigma} A_{\sigma , \nu} e^{iS_{\sigma , \nu}/\hbar} \chi_{\sigma} , \quad 0<z<d , \\
\Psi_{R} = A_{\mathit{tr}} e^{ik_{0}z} \chi_{\mathit{tr}} , \quad z>d . \nonumber 
\end{eqnarray}
Here $ A_{\mathit{inc}} $, $ A_{\mathit{ref}} $ and $ A_{\mathit{tr}} $ are amplitudes of incident, reflected and transmitted electron waves, correspondingly, and the wavenumber $ k_{E} $ is related to the energy via $ k_{E} = \sqrt{2mE/\hbar^{2}} $, $ S_{\sigma ,\nu} = \hbar k_{\sigma ,\nu} l $ with $ k_{\sigma ,\nu} = \nu q + \sigma q_{SO} $ where $ \nu = \pm 1 $ denotes the direction of the velocity and $ q = \sqrt{2m(E-V)/\hbar^{2} } $ (see Eq.(\ref{q})).

The solution in the central part depends on the form of the potential $ V(\mathbf{r} ) $. The electrostatic field in the SAM must arise from all electrons and nuclei that comprise the dsDNA molecule and its neighbours. The calculation corresponding potentials is a complicated problem and here we use only general properties of the potential (its helical symmetry) with some assumption. The double helix structure of DNA molecule contains two grooves -- major and minor; the major groove being wider then minor one. Many proteins which bind to DNA do so through wide groove. Therefore, it can be supposed that the potential minimum value corresponds to the helical trajectory $ \mathbf{r}_{0}(\tau) $ (\ref{cylpar}) with $ \rho = R $ and $ \varphi = \varphi_{0} $ which passes through the major groove. In what follows, we assume that inequality $ E - V > 0 $ takes place for trajectories which are close to the $ \mathbf{r}_{0}(\tau) $ only. For other rays $ E - V < 0 $, the quantity $ q $ becomes imeginary and the wavefunction is exponentially decaing within such trajectories. In other words, suppose that an electron passes through the SAM along the major grooves of DNA molecules as a quasifree particle.

In the functions (\ref{gensol}), the coefficients $ A $ and the components of spinors $ \chi_{\mathit{ref}} $ and $ \chi_{\mathit{tr}} $ have to be found from the boundary conditions at $ z = 0 $, $ z = d $. The wavefunction and its derivatives have to be continuous, so  
\[ \Psi_{L}(z=0) = \Psi_{C}(l=0) , \quad \bm{\nabla} \Psi_{L}(z=0) = \bm{\nabla} \Psi_{C}(l=0) , \] 
\[ \Psi_{C}(l=\mathcal{L}) = \Psi_{R}(z=d) , \quad \bm{\nabla} \Psi_{C}(l=\mathcal{L}) = \bm{\nabla} \Psi_{R}(z=d) ,\]
where $ \mathcal{L} = \left( a(R)/b \right)d $. These conditions give equations for the coefficients. Let $ A_{\mathit{inc}} = 1 $, $ \chi_{\mathit{inc}} = \left( e^{-i\beta /2} \cos (\delta /2) \quad e^{i\beta/2} \sin (\delta /2) \right)^{T} $ with $ \delta = \pi/2 - \vartheta $, and $ u_{\uparrow} $ and $ u_{\downarrow} $ are the components of $ \chi_{\mathit{tr}} $. Then one can eliminate the coefficients $ A_{\sigma ,\nu} $ and $ A_{\mathit{ref}} $ from the equations and solve for unknown components. In this case $ |\Psi_{\mathit{tr}} |^{2} = |A_{\mathit{tr}}|^{2} \left( |u_{\uparrow}|^{2} + |u_{\downarrow}|^{2} \right) = |A_{\mathit{tr}}|^{2}\equiv T $ gives the final TC value 
\begin{equation}
\label{T}
T = \frac{1}{|D|^{2}} \simeq \frac{4k_{E}^{2}q^{2}}{4k_{E}^{2}q^{2} + (k_{E}^{2} - q^{2})^{2} \sin ^{2} q\mathit{L } }
\end{equation}
with $ D = \cos q\mathcal{L} - i \left[ \left( k_{E}^{2} + q^{2} - q_{SO}^{2} /2k_{E}q\right) \right]\sin q\mathcal{L} $ and 
\begin{equation}
\label{uu}
|u_{\uparrow}|^{2} = \cos ^{2}\frac{\delta}{2} + \Delta u , \quad |u_{\downarrow}|^{2} = \sin ^{2}\frac{\delta}{2} - \Delta u
\end{equation}
where the shift
\begin{equation}
\label{Fi}
\Delta u = \frac{1}{2}\cos \vartheta \sin \gamma \sin \left( \varphi_{0} + \phi - \beta \right) \sin 2q_{SO}\mathcal{L} - 
\end{equation}
\[ -\left[ \sin \vartheta \sin ^{2}\gamma + \frac{1}{2}\cos \vartheta \sin 2\gamma \cos \left( \varphi_{0} + \phi - \beta \right)  \right] \sin ^{2}q_{SO}\mathcal{L} . \]
Therefore, at $ q_{SO}^{2} \ll k_{0}^{2} $ the expression (\ref{T}) coinsides with the well known formula for a TC of a quantum particle over a potential barrier. The SP factor  
\begin{equation}
\label{P_z}
P = \frac{T_{\uparrow} - T_{\downarrow}}{T} = |u_{\uparrow}|^{2} - |u_{\downarrow}|^{2} = \sin \vartheta + 2\Delta u (\mathcal{L})
\end{equation}
of transmitting electrons is energy independent and as a function of $ \mathcal{L} $ oscillates with the period $ \pi/q_{SO} $.

\emph{Conclusion}. Above we calculated TC for the model situation and obtained the formulas which should describe the experiments \cite{PRL06,Sci11}. It is interesting to compare them with developed theory in which the parameter (\ref{qSO}) is crucial. Its exact value for DNA molecules is unknown, but reasonable estimation can be done if to use the  Ref.\onlinecite{PRB12} and put $ q_{SO} = (m/\hbar^{2}) \alpha $, where $ \alpha = (1.87 \div 2.35)  $ meV nm \cite{PRB12,14}. Therefore the value $ \alpha \approx 2  $ meV nm should correspond to the moderate SOI. Using now for (\ref{P_z}) the width $ d = \Delta z N $ with $ N $ being the number of base pairs in dsDNA and $ \Delta z = 3.4 $ \AA{} one comes to the values $ q_{SO}\Delta z = 9\cdot 10^{-3} $. It is seen that at relevant values of the parameters the product $ q_{SO}\mathcal{L} \ll 1 $, what gives for (\ref{P_z}) linear dependence on ``optic'' length: $ P \approx P_{0} + P_{1} N $, which is evidently observed \cite{Sci11}. Because $ P_{0} = \sin \vartheta $, $ P_{1} \sim q_{SO} \Delta z $, and its sign is mainly defined by the sign of the derivative $ \partial V/\partial \rho $. And it is easy to estimate that, e.g., at $ \vartheta = 0 $ and $ N = 50 $ one has that $ P \leq 0.45 $ what is also confirmed by experiment. Thus, theoretical results are in a good qualitative and quantitative agreement with avaliable experimental data \cite{Sci11}.

This investigation was carried out in the framework of Special Programm of Fundamental Researches of NAS of Ukraine.


\begin{thebibliography}{99}
\bibitem{tbappr} 
The tight-binding approximation is considered in Ref.\onlinecite{PRB12}. For the model with the nearest neighbour hopping approximation, the Hamiltonian can be diagonalized: $ H = \sum_{k,\sigma}E_{\sigma}(k) a^{\dagger}_{k,\sigma}a_{k,\sigma} $. Here $ \sigma = \pm 1 $ is the spin and (in the notations of Ref.\onlinecite{PRB12}) $ E_{\pm}(k) = U_{0} + 2V_{\mathit{eff}} \cos (k \pm q_{SO})a $ with $ a $ being the lattice constant, $ V_{\mathit{eff}} = V\sqrt{1 + (|W|/V)^{2}} $ and $ \tan q_{SO}a = |W|/V $. The parameters $ V $ and $ W $ describe hopping without and with spin-flip due to the SOI.
\end{thebibliography}
\end{document}